\documentclass[
  aps,
  prb,                    
  reprint,                
  superscriptaddress,
  amsmath,
  amssymb,
  nofootinbib
]{revtex4-2}

\usepackage[T1]{fontenc}
\usepackage[utf8]{inputenc} 
\usepackage{lmodern}
\usepackage{microtype}
\usepackage{graphicx}
\usepackage{dcolumn}
\usepackage{bm}
\usepackage{siunitx}
\usepackage{booktabs}
\usepackage{xcolor}

\newif\ifmarkcorrections
\markcorrectionstrue

\usepackage{hyperref}

\hypersetup{
  colorlinks=true,
  linkcolor=blue,
  citecolor=blue,
  urlcolor=blue
}

\begin{document}

\title{Acoustic bound states in the continuum in a labyrinthine meta-atom}

\author{Aleksandra S. Zheleznova}
\thanks{A.S. Zheleznova, I.A. Timankova and ZiWei Ye contributed equally to this work.}
\affiliation{School of Physics and Engineering, ITMO University, St.\ Petersburg, Russia}

\author{Iuliia A. Timankova}
\thanks{A.S. Zheleznova, I.A. Timankova and ZiWei Ye contributed equally to this work.}
\affiliation{School of Physics and Engineering, ITMO University, St.\ Petersburg, Russia}

\author{ZiWei Ye}
\thanks{A.S. Zheleznova, I.A. Timankova and ZiWei Ye contributed equally to this work.}
\affiliation{Qingdao Innovation and Development Center, Harbin Engineering University, Qingdao, China}

\author{Yunmeng Wang}
\affiliation{College of Underwater Acoustic
Engineering, Harbin Engineering University, Harbin, China}

\author{Jihao Tang}
\affiliation{College of Underwater Acoustic
Engineering, Harbin Engineering University, Harbin, China}

\author{Mikhail V. Kuzmin}
\affiliation{School of Physics and Engineering, ITMO University, St.\ Petersburg, Russia}

\author{Nikolay S. Solodovchenko}
\affiliation{School of Physics and Engineering, ITMO University, St.\ Petersburg, Russia}

\author{Mihail I. Petrov}
\affiliation{School of Physics and Engineering, ITMO University, St.\ Petersburg, Russia}

\author{Andrey A. Bogdanov}
\email{a.bogdanov@metalab.ifmo.ru}
\affiliation{Qingdao Innovation and Development Center, Harbin Engineering University, Qingdao, China}
\affiliation{School of Physics and Engineering, ITMO University, St.\ Petersburg, Russia}

\author{Pavel S. Pankin}
\email{pankin@itmo.ru}
\affiliation{School of Physics and Engineering, ITMO University, St.\ Petersburg, Russia}

\author{Yong Li}
\affiliation{School of Physics and Engineering, ITMO University, St.\ Petersburg, Russia}
\affiliation{Institute of Acoustics, School of Physics Science and Engineering, Tongji University, Shanghai 200092, China}
\date{\today}

\begin{abstract}
In this Letter, we theoretically and experimentally study acoustic bound states in the continuum (BICs) emerging in labyrinth meta-atom embedded in an acoustic waveguide.
A symmetry-protected BIC is supported in the lossless limit when the meta-atom is positioned on the waveguide axis. Breaking the symmetry by shifting the meta-atom away from the center of the waveguide transforms the BIC into a radiative quasi-BIC, manifested by the emergence of a narrow resonant feature in the transmittance spectrum. The proposed concept is experimentally tested using a 3D-printed labyrinthine meta-atom placed inside an impedance tube. The measured transmittance spectra are consistent with numerical simulations and support the BIC-to-quasi-BIC transition induced by symmetry breaking in the presence of thermoviscous losses. The resonance frequency, contrast, and linewidth can be controlled by varying the meta-atom position. The observed behavior is interpreted within a two-mode temporal coupled-mode theory (TCMT), providing an analytical description of the resonant response. The ability to tune radiative leakage in the presence of absorption makes the proposed acoustic BIC platform promising for the development of tunable absorbers, sensors, and modulators.

\end{abstract}


\maketitle


Bound states in the continuum (BICs) are spatially localized eigenstates whose frequencies lie inside a spectrum of propagating waves while remaining completely decoupled from the radiation channels.
In the ideal case, the radiative linewidth $2\gamma_r$ of such a state vanishes and its radiative quality factor (Q-factor) $Q_r=\omega_0/2\gamma_r$ diverges.
Initially introduced in quantum mechanics, the BIC concept has subsequently become an important paradigm in wave physics, particularly in photonics~\cite{Hsu2016,wang2024optical}.
In realistic systems, a perturbation that opens a radiation channel transforms an ideal BIC into a quasi-BIC with a finite but potentially very high and tunable Q-factor.


\begin{figure*}[t]
  \centering
  \includegraphics[width=1\textwidth]{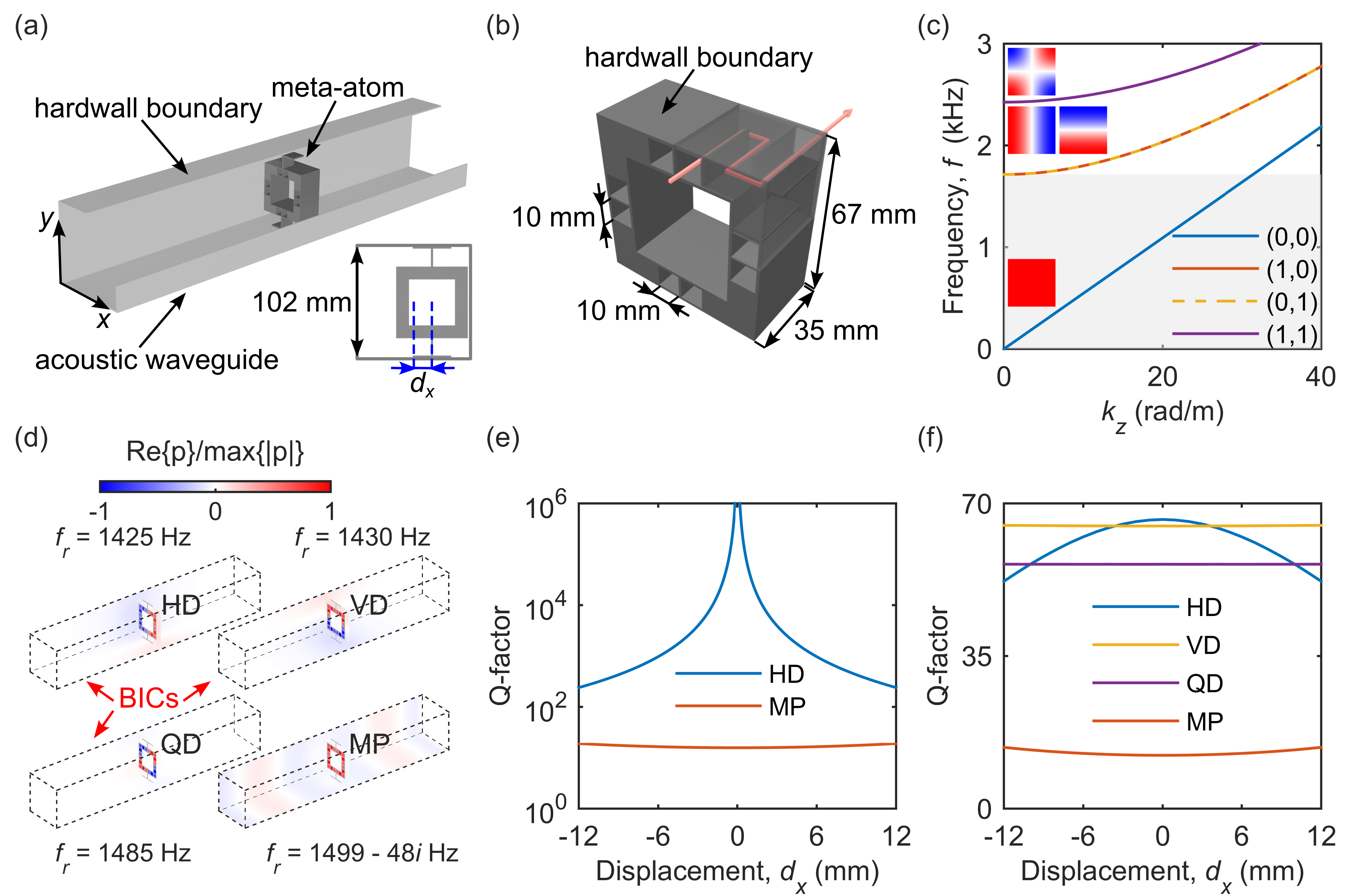}
  \caption{(a) The labyrinthine meta-atom placed in the acoustic waveguide.
The inset shows the displacement $d_x$.
  (b) The labyrinthine meta-atom.
The pink arrow shows the sound propagation path.
  (c) Dispersion curves for waveguide modes.
The insets show the corresponding pressure field profiles $\mathrm{Re}\{p\}/\max\{|p|\}$ .
  Gray shading marks the single-mode range below the first higher-order cutoff.  
  (d) Meta-atom eigenmode pressure field profiles $\mathrm{Re}\{p\}/\max\{|p|\}$ and corresponding eigenfrequencies $f_r=\omega_r/2\pi$ without thermoviscous losses at $d_x=0$. (e,f) Q-factor vs displacement in the lossless (e) and lossy cases (f).
HD, VD, QD and MP denote the vertical dipole, horizontal dipole, quadrupole and monopole modes, respectively.
  The VD and QD modes have $Q_r\to\infty$ throughout the displacement range in (e).}
  \label{Fig_1}
\end{figure*}

The BIC concept has been successfully transferred to acoustics, where different types of trapped states have been demonstrated in open acoustic resonators and coupled waveguide--resonator systems~\cite{Huang2021Sound,Huang2022Framework,Jia2023ReducedSymmetry}.
Symmetry-protected and accidental acoustic BICs, including those arising from Friedrich--Wintgen interference, can be obtained by controlling the symmetry and coupling between localized resonant modes and propagating waves in scattering channels~\cite{Huang2021Sound,Huang2022Framework,Jia2023ReducedSymmetry,Yin2024Rotating}.
Genuine BICs can also exist in finite three-dimensional compact solid resonators coupled to an acoustic environment, where symmetry breaking converts them into high-$Q$ quasi-BICs observable as narrow Fano resonances in the scattering spectrum~\cite{Deriy2022Compact}.
More recently, acoustic BICs have been directly observed in an open two-dimensional waveguide geometry, allowing experimental characterization not only through scattering spectra but also through the spatial distribution of the localized acoustic field~\cite{MartiSabate2024Observation}.
Merged Fabry--Perot BICs in underwater phononic crystals further extend the range of acoustic configurations supporting BICs~\cite{Yang2024Merged}.
Phononic and elastic systems also support BIC lines~\cite{BICLines2025}, multibranch states~\cite{Multibranch2024}, valley edge states embedded in the continuum~\cite{Yin2024Valley}, and hybridized states involving different elastic polarizations~\cite{Liu2025Elastic}.

Particular attention has recently been devoted to geometrically tunable quasi-BICs and coupled acoustic resonators.
Accidental BICs can be controlled through rotation of internal obstacles in waveguide--resonator systems~\cite{Yin2024Rotating}, whereas coupled Helmholtz resonators provide a simple platform in which the transition to quasi-BIC states can be controlled through coupling and symmetry breaking~\cite{Krasikova2024Helmholtz}.
The latter configuration allows the interplay between the radiative decay rate $\gamma_r$ and the absorptive decay rate $\gamma_a$ to be investigated. Recent topology-optimization approaches have also enabled the systematic design of BIC-based high-$Q$ acoustic structures with enhanced field localization~\cite{Li2025Topology}, while collective configurations have been proposed to broaden the bandwidth of acoustic emission enhancement~\cite{Huang2025Collective} and multiple quasi-BICs have been used for multiband sensing~\cite{Jiao2025Multiband}.

The strong acoustic-field localization associated with quasi-BICs has stimulated a range of practical applications.
Quasi-BICs have been employed to enhance acoustic emission through the Purcell effect~\cite{Huang2024Purcell}, to realize high-sensitivity and frequency-selective acoustic sensing~\cite{Song2025Sensor,Jiao2025Multiband}, and to improve acoustic energy harvesting~\cite{Kronowetter2025Harvesting}.
These developments and other applications of acoustic and elastic BICs are summarized in the recent perspective by He \textit{et al.}~\cite{He2025Review}.
In this Letter, we theoretically investigate and experimentally implement a platform for acoustic BICs based on a labyrinthine meta-atom placed in a waveguide. 

The structure under study is shown in Fig.~\ref{Fig_1}(a).
The acoustic waveguide has rigid walls and a square cross section with side length $a=102$~mm.
The inset of Fig.~\ref{Fig_1}(a) defines $d_x$ as the displacement of the meta-atom from the waveguide center along $x$, perpendicular to the propagation direction $z$.
The dispersion law $\omega_{m,n}=c\sqrt{(m\pi/a)^2+(n\pi/a)^2+k_z^2}$ and mode profiles for the lowest-frequency waveguide modes are shown in Fig.~\ref{Fig_1}(c).
The operating regime lies below the first higher-order cutoff $f_c=c/2a\approx 1685$~Hz for $c=343.79$~m/s, where only the fundamental waveguide mode $(m,n)=(0,0)$ propagates in each direction.
Here, $m,n$ are nonnegative transverse mode indices and $k_z$ is the propagation constant.

The eigenvalue problem for the meta-atom placed in the waveguide was solved by the finite-element method (FEM) in COMSOL Multiphysics.
The meta-atom's  labyrinthine structure is designed to provide an extended propagation path for sound waves and, consequently, a reduced effective phase velocity, see Fig.~\ref{Fig_1}(b)~\cite{wu2021acoustic,timankova2026experimental}.
This coiling of the propagation path brings the meta-atom eigenmodes into the operating frequency range.
In homogeneous air in the lossless model, the acoustic pressure satisfies the Helmholtz equation $\nabla^2p+(\omega/c)^2p=0$, with $\partial_n p=0$ at rigid boundaries and outgoing-wave conditions at the waveguide ends.
The eigenmode profiles and corresponding eigenfrequencies $\omega_r=\omega_0-i\gamma_r$ in the operating frequency range are shown in Fig.~\ref{Fig_1}(d).
We use the time convention $e^{-i\omega t}$, then $\gamma_r\geq0$ corresponds to an amplitude decay rate.
In contrast to the monopole mode, which radiates into the waveguide, dipole and quadrupole modes are localized near the meta-atom at zero displacement and have purely real eigenfrequencies.

\begin{figure*}[t]
\centering

\small
\renewcommand{\arraystretch}{1.15}
\setlength{\tabcolsep}{4pt}

\newcommand{\modewidth}{0.085\textwidth}

\begin{tabular}{|c|c|c|c|c|c|c|}
\hline

Mode
&
\begin{tabular}{c}
Eigenfrequency\\
(Hz)
\end{tabular}
&
$Q$-factor
&
\raisebox{-0.5\height}{
\includegraphics[width=4cm]{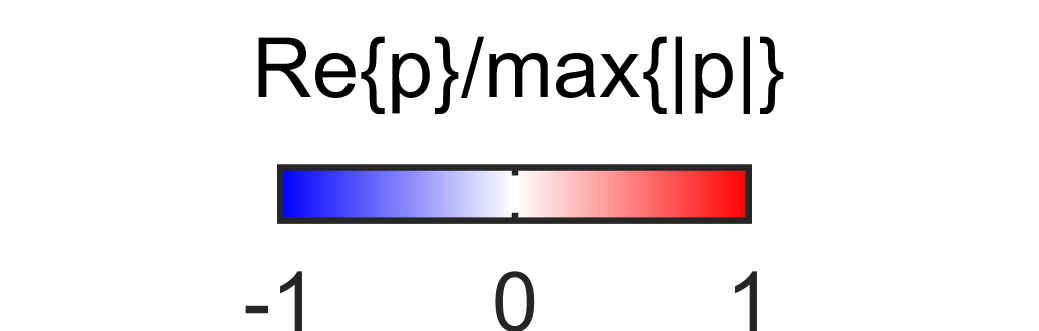}
}
&
$\sigma(xz)$
&
$\sigma(yz)$
&
IrRep
\\
\hline

\multicolumn{7}{|c|}{\textbf{Waveguide mode}}
\\
\hline

Fundamental
&
Single-mode  range  $\left[0,\;1715\right]$
&
$\infty$
&
\raisebox{-0.5\height}{
\includegraphics[width=\modewidth]{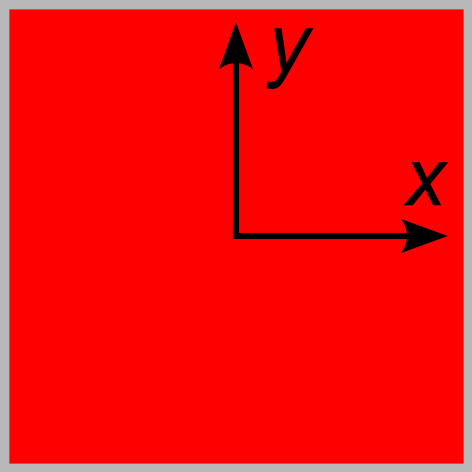}
}
&
$+1$
&
$+1$
&
$A_1$
\\
\hline

\multicolumn{7}{|c|}{\textbf{Meta-atom modes}}
\\
\hline

Horizontal Dipole (HD)
&
$1425$
&
$\infty$
&
\raisebox{-0.5\height}{
\includegraphics[width=\modewidth]{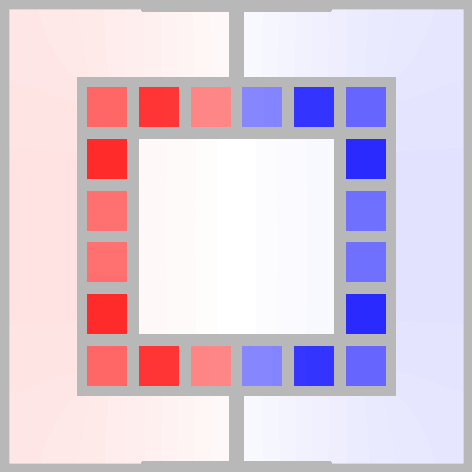}
}
&
$+1$
&
$-1$
&
$B_1$
\\
\hline

Vertical Dipole (VD)
&
$1430$
&
$\infty$
&
\raisebox{-0.5\height}{
\includegraphics[width=\modewidth]{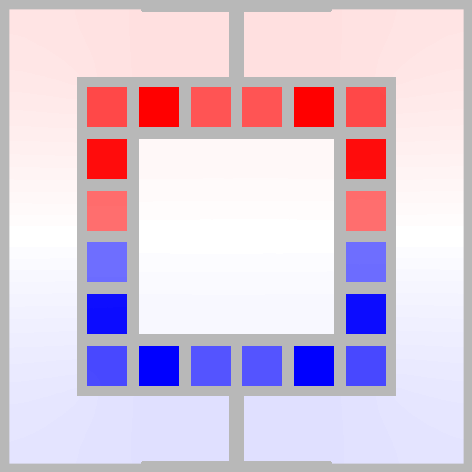}
}
&
$-1$
&
$+1$
&
$B_2$
\\
\hline

Quadrupole (QD)
&
$1485$
&
$\infty$
&
\raisebox{-0.5\height}{
\includegraphics[width=\modewidth]{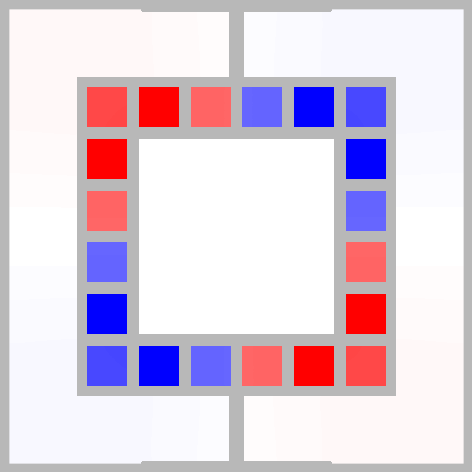}
}
&
$-1$
&
$-1$
&
$A_2$
\\
\hline

Monopole (MP)
&
$1499 - 48i$
&
$15.6$
&
\raisebox{-0.5\height}{
\includegraphics[width=\modewidth]{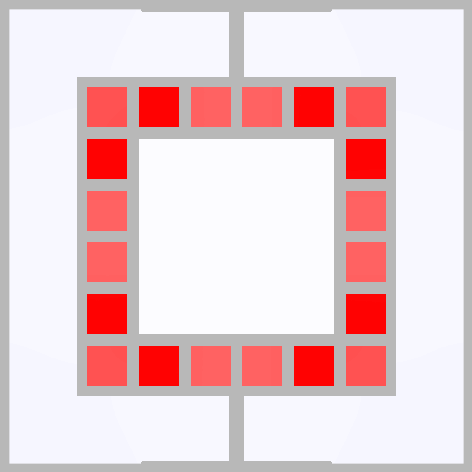}
}
&
$+1$
&
$+1$
&
$A_1$
\\
\hline

\end{tabular}

\caption{Eigenfrequencies, quality factors, pressure distributions,
symmetry properties, and irreducible representations of the eigenmodes of the centered structure, classified by the $C_{2v}$ point group in the lossless case.}

\label{Fig_2}

\end{figure*}

Figures ~\ref{Fig_1}(e) and ~\ref{Fig_1}(f) show the dependence of the Q-factor on the displacement $d_x$.
Figure~\ref{Fig_1}(e) corresponds to the case without thermoviscous losses.
The radiative Q-factor of the horizontal dipole diverges in the symmetric case, when the meta-atom is not shifted.
At the same time, the radiative Q-factors of the vertical dipole and quadrupole remain infinite throughout the displacement range considered.
This behavior can be explained by examining the symmetries of the eigenmodes and propagating waveguide modes, see Fig.~\ref{Fig_2}.
All four meta-atom modes belong to different irreducible representations of the $C_{2v}$ point group.
At the same time, the monopole and waveguide modes belong to the same irreducible representation.
Thus, the dipole and quadrupole modes in the symmetric case are orthogonal to the only propagating waveguide mode, and consequently cannot radiate~\cite{sadrieva2019multipolar}.

Translation along the $x$ axis preserves reflection symmetry in the $xz$ plane.
The vertical dipole and quadrupole are odd under this reflection, whereas the propagating mode is even, so their orthogonality is preserved throughout the displacement range for $0<f<f_c$, where only the fundamental mode propagates.
At the same time, reflection symmetry in the $yz$ plane is broken, which permits coupling between the horizontal dipole and the waveguide mode and decreases the radiative Q-factor.
For small displacements, a nonzero linear radiation-coupling term gives $\gamma_r\propto d_x^2$ and hence $Q_r\propto d_x^{-2}$.
In Fig.~\ref{Fig_1}(f), the total Q-factor is limited by absorption (thermoviscous) losses $Q=\omega_0/2(\gamma_r+\gamma_a)$, or $Q^{-1}=Q_r^{-1}+Q_a^{-1}$ with $Q_a=\omega_0/2\gamma_a$.
At the symmetry-protected point, $\gamma_r=0$ and $Q=Q_a$ remains finite.
In both cases, the Q-factor for the monopole mode is finite, since it has the same parity as the fundamental waveguide mode, see Fig.~\ref{Fig_2}.

\begin{figure*}[t]
  \centering
  \includegraphics[width=1\textwidth]{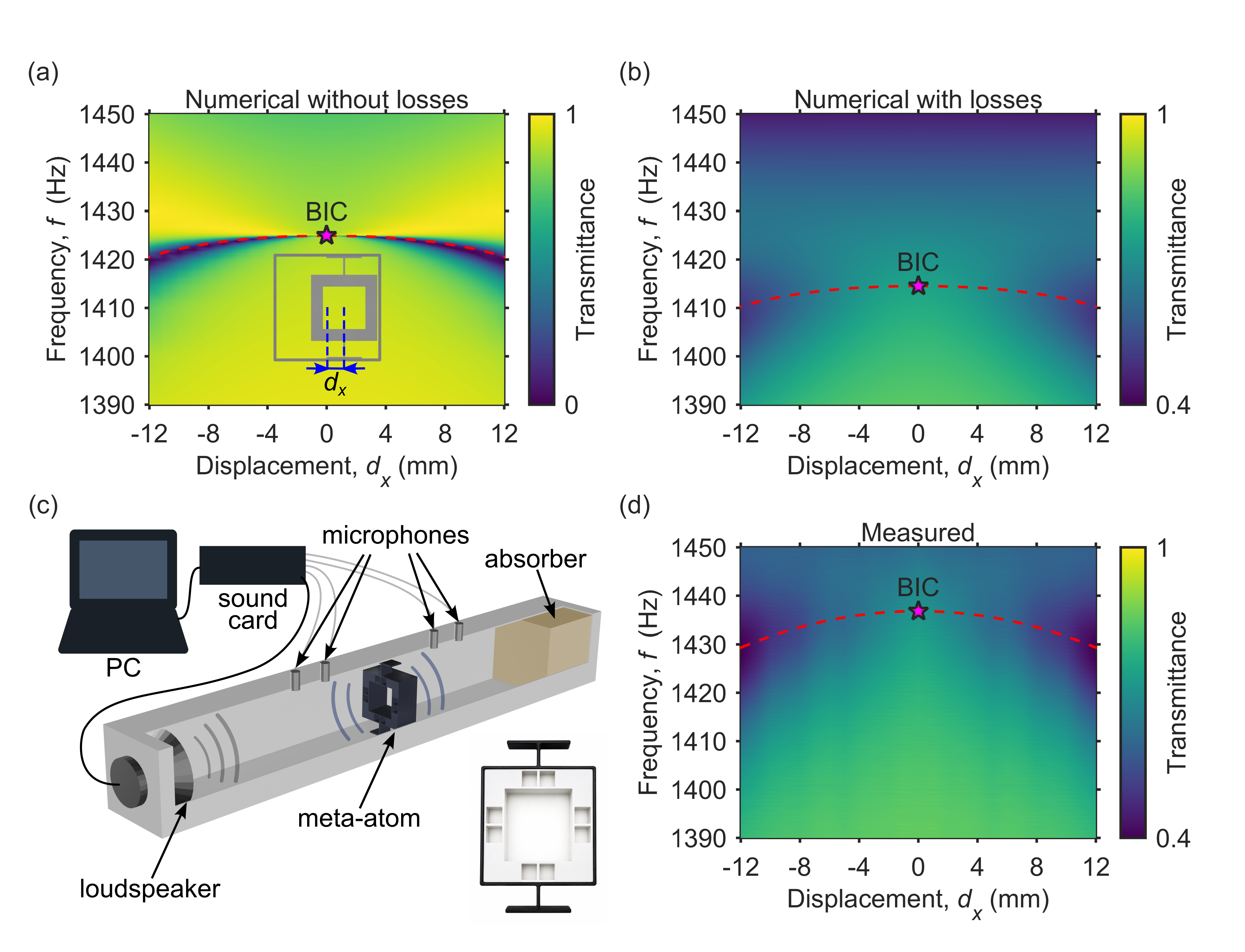}
  \caption{The numerically calculated transmittance spectra in the lossless (a) and lossy cases (b).
The stars mark the BIC points.
The red dashed line shows the resonance frequency $f_0=\omega_0/2\pi$.
(c) The experimental setup.
The inset shows the printed meta-atom.
(d) The measured transmittance spectra.
The red dashed line shows the minimum of the resonant transmittance, approximated by a polynomial through the central part.
The star marks the resonant dip vanishing point.}
  \label{Fig_3}
\end{figure*}

The waveguide transmittance spectra were calculated in COMSOL Multiphysics with scattering boundary conditions at the waveguide ends.
The results without and with thermoviscous losses are shown in Fig.~\ref{Fig_3}(a,b), respectively.
In COMSOL Multiphysics, losses were taken into account via the Thermoviscous Boundary Layer Impedance (TBLI); see the physical parameters in the Supplemental Material.
The asymmetric Fano resonance corresponding to the horizontal dipole quasi-BIC mode can be observed on the background formed by the broadband monopole resonance.
The radiative linewidth vanishes as the symmetric configuration is approached, which corresponds to the BIC limit, see Fig.~\ref{Fig_3}(a).
At exact symmetry, the dipole mode cannot be excited by the incident plane wave and leaves no resonance in the transmission spectrum.
Remaining decoupled from the propagating waveguide mode at all displacements $d_x$ considered, the vertical dipole and quadrupole cannot be excited by the incident plane wave and do not produce resonant lines in the spectra.
Thermoviscous losses broaden and shift the horizontal dipole resonance and reduce its amplitude while the linewidth remains finite, see Fig.~\ref{Fig_3}(b).
In contrast to the lossless case, for which a resonant transmission zero always exists  for nonzero displacement, in the lossy case, the resonant dip becomes more pronounced as the displacement magnitude $|d_x|$ increases over the range considered.

For the experimental implementation, the labyrinthine meta-atom was fabricated via fused deposition modeling (FDM 3D printing) using polylactic acid (PLA).
The schematic of the experimental setup is shown in Fig.~\ref{Fig_3}(c).
The acoustic waveguide consists of a metallic  impedance tube with a square cross section with side length $a=102$~mm incorporating four microphones, a loudspeaker that produces the incident sound wave, and an absorbing termination.
Two microphones on each side of the sample resolve the forward and backward plane waves. From these amplitudes we obtained the transmission coefficient $S_{21}$ and power transmittance $T=|S_{21}|^2$, see Supplemental Material.
Figure~\ref{Fig_3}(d) shows a color map of the measured transmittance as a function of the meta-atom displacement $d_x$.
The resonant dip disappears at zero offset $d_x = 0$ mm, but absorption keeps the mode linewidth finite, in agreement with the numerical spectra.


The observed spectral features can be interpreted in the framework of the temporal coupled-mode theory (TCMT)~\cite{fan2003temporal}.
Since the dipole resonance appears on the background of the monopole resonance, we use the two-mode TCMT equations of Ref.~\cite{suh2004temporal}

\begin{equation}
\label{TCMT_1}
\frac{d\mathbf{a}}{dt}
=
-
\left(
i\widehat{\Omega}
+\widehat{\Gamma}_r
+\widehat{\Gamma}_a
\right)
\mathbf{a}
+
\widehat{D}^{T}\mathbf{s}^{(+)},
\end{equation}

\begin{equation}
\label{TCMT_2}
\mathbf{s}^{(-)}
=
\widehat{C}\mathbf{s}^{(+)}
+
\widehat{D}\mathbf{a},
\end{equation}
where
\begin{equation}
\label{a_and_s}
\mathbf{a}
=
\begin{pmatrix}
a_1\\
a_2
\end{pmatrix},
\qquad
\mathbf{s}^{(+)}
=
\begin{pmatrix}
s_1^{(+)}\\
s_2^{(+)}
\end{pmatrix},
\qquad
\mathbf{s}^{(-)}
=
\begin{pmatrix}
s_1^{(-)}\\
s_2^{(-)}
\end{pmatrix}.
\end{equation}
Here $a_n$ is the $n$th resonant mode amplitude, with $n=1$ for the monopole and $n=2$ for the horizontal dipole.
The amplitudes $s_m^{(\pm)}$ describe the incoming $(+)$ and outgoing $(-)$ waveguide modes at the left ($m=1$) and right ($m=2$) scattering channels along the waveguide axis $z$.
We use energy-normalized resonant amplitudes and power-normalized scattering-channel amplitudes, so $|a_n|^2$ is an energy and $|s_m^{(\pm)}|^2$ is a power.

The direct (nonresonant) scattering matrix $\widehat{C}$ is approximated as lossless and frequency independent over the operating frequency range.
With scattering-channel reference phases chosen to make its diagonal entries equal, it is written as~\cite{maksimov2025temporal}
\begin{equation}
\label{C_matrix}
\widehat{C}
=
\begin{pmatrix}
\rho & i\tau \\
i\tau & \rho
\end{pmatrix}, \;
\rho^2+\tau^2=1.
\end{equation}
The nonresonant reflection and transmission amplitudes are $\rho$ and $i\tau$, respectively, where $\rho$ and $\tau \in [-1, \; 1]$ are real-valued parameters. 
The matrices of resonant frequencies $\widehat{\Omega}$ and absorptive decay rates $\widehat{\Gamma}_a$ are written as
\begin{equation}
\label{Omega_Gamma0}
\widehat{\Omega}
=
\begin{pmatrix}
\omega_{01} & J\\
J & \omega_{02}
\end{pmatrix}, \;
\widehat{\Gamma}_a
=
\begin{pmatrix}
\gamma_{a1} & 0\\
0 & \gamma_{a2}
\end{pmatrix},
\end{equation}
where we introduce the near-field coupling $J$ between the two resonances and neglected off-diagonal absorption.
The matrices of radiative decay rates $\widehat{\Gamma}_r$ and decoupling constants $\widehat{D}$ are written as
\begin{equation}
\label{Gamma_D}
\widehat{\Gamma}_r
=
\begin{pmatrix}
\gamma_{r1} & \kappa_{12}\\
\kappa_{21} & \gamma_{r2}
\end{pmatrix}, \;
\widehat{D}
=
\begin{pmatrix}
d_{11} & d_{12}\\
d_{21} & d_{22}
\end{pmatrix}.
\end{equation}
The resonant modes are coupled via the common scattering channels with corresponding coupling coefficients $\kappa_{12}$ and $\kappa_{21}$, as in the Friedrich--Wintgen model~\cite{sadreev2021interference}.
The diagonal terms in the matrices $\widehat{\Omega}$, $\widehat{\Gamma}_a$ and $\widehat{\Gamma}_r$ correspond to the decoupled modes.

\begin{figure*}[t]
  \centering
\includegraphics[width=\linewidth]{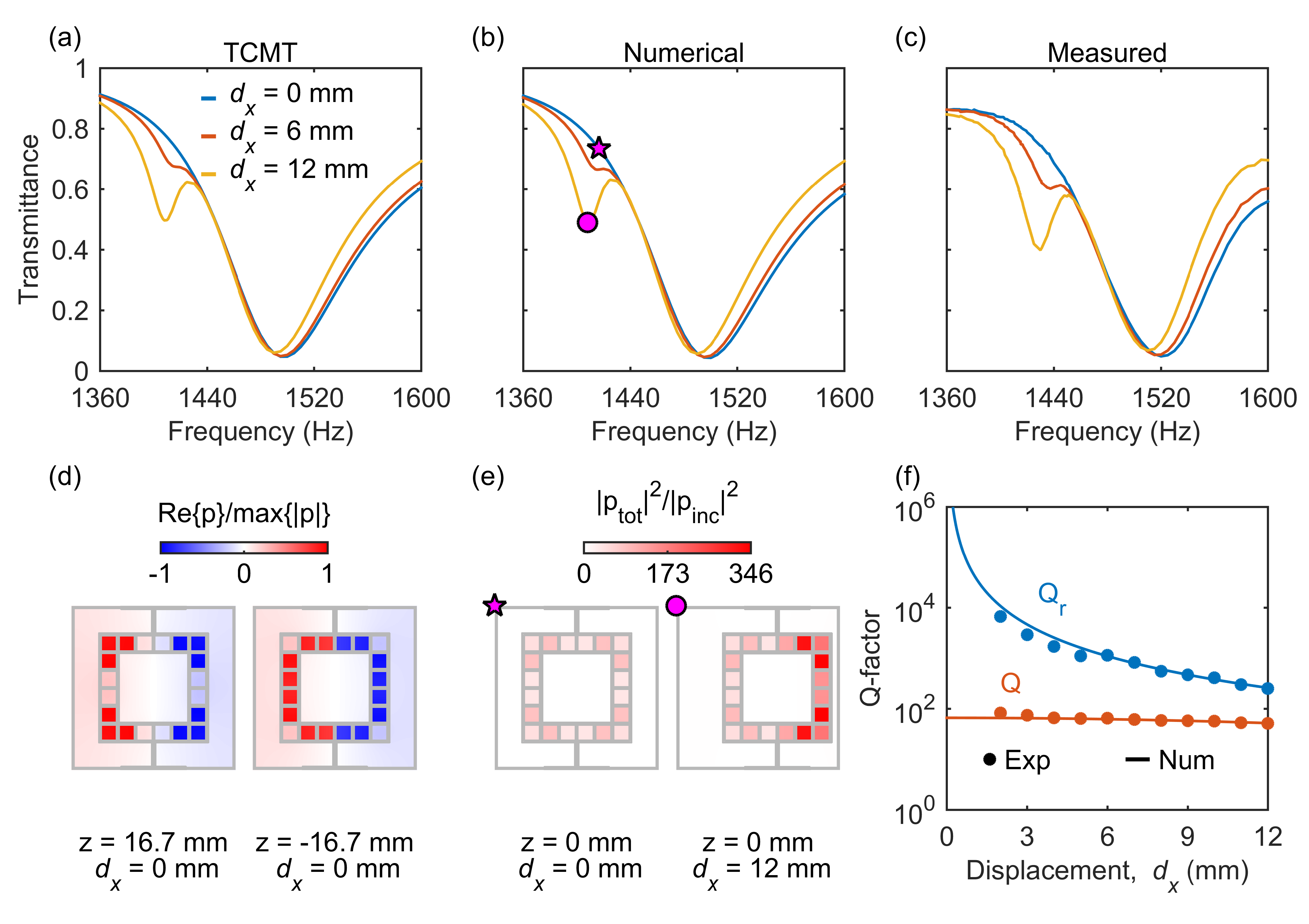}
  \caption{The transmittance spectra obtained by TCMT (a), numerically calculated by FEM (b), and measured in the impedance tube (c).
(d) Horizontal dipole eigenmode pressure field profiles $\mathrm{Re}\{p\}/\max\{|p|\}$ near the front and back walls.
(e) Numerically calculated normalized squared total pressure amplitude $|p|^{2}/|p_{inc}|^{2}$ at the resonant frequencies denoted with markers in (b). Here $p_{inc}$ is the incident waveguide mode pressure field amplitude.
(f) Radiative ($Q_r$, blue) and total ($Q$, orange) quality factors of the horizontal dipole. Lines show FEM eigenmode results. Points are obtained from Fano fits to experimental extinction, with $Q_r$ estimated under the assumption of equal radiation into the two ports, see Supplemental Material.
}
  \label{Fig_4}
\end{figure*}
Since the meta-atom is not symmetric with respect to the $xy$ plane [Fig.~\ref{Fig_1}(b)], its monopole and dipole modes can couple asymmetrically to the two scattering channels.
Their radiation coupling vectors can be parametrized as~\cite{maksimov2025temporal}

\begin{equation}
\label{d_vector}
\begin{pmatrix}
d_{1n}\\
d_{2n}
\end{pmatrix}
=
\sqrt{
\frac{\gamma_{rn}}
{1+\rho}
}
\begin{pmatrix}
\tau \cos{\alpha_n}-i(1+\rho)\sin{\alpha_n}\\
\tau \sin{\alpha_n}-i(1+\rho)\cos{\alpha_n}
\end{pmatrix},
\end{equation}
where $\alpha_n\in[-\pi/2,\pi/2]$ is an asymmetry parameter.
For a nonabsorbing direct process, energy conservation gives
\begin{equation}
\label{DD_Gamma}
\widehat{D}^{\dagger}\widehat{D}
=
2\widehat{\Gamma}_r.
\end{equation}
%
Equation~\eqref{d_vector} then gives
\begin{equation}
\kappa_{12}=\kappa_{21}=\sqrt{\gamma_{r1}\gamma_{r2}}\cos(\alpha_1-\alpha_2).
\end{equation}
Thus the off-diagonal radiative coupling is fixed by the radiation vectors and is not an independent fitting parameter. 

From Eq.~\eqref{TCMT_1}, under a steady-state excitation with time dependence $e^{-i\omega t}$, the vector of the resonant mode amplitudes becomes
\begin{equation}
\label{amplitude}
\mathbf{a}
=
\left[i\left(\widehat{\Omega}-\omega\widehat{I}\right)
+
\widehat{\Gamma}_r
+
\widehat{\Gamma}_a\right]^{-1}\widehat{D}^{T}
\mathbf{s}^{(+)}.
\end{equation}
Using Eq.~\eqref{amplitude} for incidence from the first scattering channel, i.e., $s_1^{(+)}=1$ and $s_2^{(+)}=0$, the transmittance $T$ can be found directly from Eq.~\eqref{TCMT_2} as
\begin{equation}
\label{T}
T(\omega)
\equiv
\left|
\frac{s_2^{(-)}}{s_1^{(+)}}
\right|^2
=
\left|
C_{21}+d_{21}a_1+d_{22}a_2
\right|^2.
\end{equation}%
Equation~\eqref{T} describes interference between three contributions, namely the directly transmitted wave and the waves radiated by the monopole and dipole modes.

The spectra obtained by TCMT, FEM, and measurements are compared in Fig.~\ref{Fig_4}(a--c).
For fitting the TCMT parameters, we exploited the symmetry properties of the eigenmodes shown in Fig.~\ref{Fig_2}. At $d_x=0$, the horizontal dipole is BIC, $\gamma_{r2}(0)=0$.
The near-field coupling between the monopole and horizontal dipole modes also vanishes, $J(0)=0$, since these modes belong to different irreducible representations of the $C_{2v}$ point group and have opposite parities with respect to reflection in the $yz$ plane. Under the transformation $d_x\rightarrow-d_x$, the near-field coupling $J$ changes sign, so its expansion contains only odd powers of $d_x$, with the leading term $J\propto d_x$. This linear dependence is allowed by the symmetry of the overlap integral between the two modes and the first-order symmetry-breaking perturbation. 
The eigenfrequencies $\omega_{0n}$, the radiative decay rates $\gamma_{rn}$, and the direct transmission parameter $\tau$ are even functions of $d_x$, with leading corrections proportional to $d_x^2$.
The asymmetry parameters $\alpha_1$ and $\alpha_2$ account for the unequal coupling of the resonant modes to the two scattering channels, arising from the lack of mirror symmetry with respect to the $xy$ plane, as illustrated by the different pressure field distributions on opposite sides of the meta-atom in Fig.~\ref{Fig_4}(d).
We held the asymmetry parameters $\alpha_1$ and $\alpha_2$ and the absorptive decay rates $\gamma_{a1}$ and $\gamma_{a2}$ constant across displacements in the fit.
These symmetry constraints were used to approximate the displacement dependence of the TCMT parameters by polynomials, whose coefficients were fitted to the numerical spectra.
The resulting dependencies are presented in the Supplemental Material.

In the symmetric case $d_x = 0$, only the broad monopole resonance dip is observed.
Breaking the symmetry allows the horizontal dipole to couple to the waveguide mode, producing a quasi-BIC resonance.
The amplitude, width, and position of the dipole quasi-BIC resonance change with increasing displacement $d_x$, demonstrating the tunability of the transmittance spectra.
The coupling between the two modes is confirmed by the field distribution in Fig.~\ref{Fig_4}(e).
At $d_x = 0$ mm, only the monopole mode is excited, whereas at $d_x = 12$ mm, the asymmetric field distribution confirms the excitation of the coupled modes.
For the experimental points in Fig.~\ref{Fig_4}(f), we fitted the normalized extinction $\sigma_{\mathrm{ext}}$ calculated from the complex transmission amplitude. The fitted dipole linewidth gives $Q$, and the line amplitude gives an estimate of $Q_r$.
The spectra in Fig.~\ref{Fig_4}(a--c), as well as the Q-factors in Fig.~\ref{Fig_4}(f), are in good qualitative agreement, which confirms the realization of the new platform for acoustic BICs based on labyrinthine meta-atoms.

In conclusion, the lossless eigenmode calculation predicts symmetry-protected BICs in a centered labyrinthine meta-atom. In the impedance-tube experiment, shifting the printed meta-atom off center produces a narrow dipole feature that disappears when the meta-atom returns to the center. This displacement dependence supports the BIC prediction in a sample with finite absorption.
The tunability of the resonant line amplitude, width, and position is shown.
The observed response is interpreted using two-mode temporal coupled-mode theory.
The eigenmode symmetry can be changed by modifying the symmetry of the printed meta-atom, while the spectral position can be adjusted by varying the effective parameters determined by the distribution of the meta-atom channels.  
The proposed platform offers a route toward acoustic
resonances with tunable coupling for absorbers, sensors, and modulators.

\begin{acknowledgments}
This work was funded by the Russian Science Foundation, Grant No.~25-79-31027 (\url{https://rscf.ru/project/25-79-31027/}).
\end{acknowledgments}


\bibliography{References}

@article{sadreev2021interference,
  author  = {Sadreev, Almas F.},
  title   = {Interference traps waves in an open system: bound states in the continuum},
  journal = {Reports on Progress in Physics},
  volume  = {84},
  number  = {5},
  pages   = {055901},
  year    = {2021},
  doi     = {10.1088/1361-6633/abefb9},
  url     = {https://doi.org/10.1088/1361-6633/abefb9}
}

@article{timankova2026experimental,
  author  = {Timankova, I. A. and Smagin, M. V. and Kuzmin, M. V. and Lutovinov, A. I. and Bogdanov, A. A. and Li, Y. and Petrov, M. I.},
  title   = {Experimental Investigation of Acoustic {Kerker} Effect in Labyrinthine Resonators},
  journal = {JETP Letters},
  volume  = {123},
  number  = {2},
  pages   = {133--139},
  year    = {2026},
  doi     = {10.1134/S0021364025609674},
  url     = {https://doi.org/10.1134/S0021364025609674}
}

@article{maksimov2025temporal,
  author  = {Maksimov, Dmitrii N. and Pankin, Pavel S. and Kim, Dong-Wook and Song, Mingzhao and Peng, Chao and Bogdanov, Andrey A.},
  title   = {Temporal coupled mode theory: From bound states in the continuum to uniguided resonances},
  journal = {Physical Review B},
  volume  = {112},
  number  = {23},
  pages   = {235303},
  year    = {2025},
  doi     = {10.1103/rq46-g378},
  url     = {https://doi.org/10.1103/rq46-g378}
}

@article{suh2004temporal,
  author  = {Suh, Wonjoo and Wang, Zheng and Fan, Shanhui},
  title   = {Temporal coupled-mode theory and the presence of non-orthogonal modes in lossless multimode cavities},
  journal = {IEEE Journal of Quantum Electronics},
  volume  = {40},
  number  = {10},
  pages   = {1511--1518},
  year    = {2004},
  doi     = {10.1109/JQE.2004.834773},
  url     = {https://doi.org/10.1109/JQE.2004.834773}
}

@article{fan2003temporal,
  author  = {Fan, Shanhui and Suh, Wonjoo and Joannopoulos, J. D.},
  title   = {Temporal coupled-mode theory for the {Fano} resonance in optical resonators},
  journal = {Journal of the Optical Society of America A},
  volume  = {20},
  number  = {3},
  pages   = {569--572},
  year    = {2003},
  doi     = {10.1364/JOSAA.20.000569},
  url     = {https://doi.org/10.1364/JOSAA.20.000569}
}

@article{sadrieva2019multipolar,
  author  = {Sadrieva, Zarina and Frizyuk, Kristina and Petrov, Mihail and Kivshar, Yuri and Bogdanov, Andrey},
  title   = {Multipolar origin of bound states in the continuum},
  journal = {Physical Review B},
  volume  = {100},
  number  = {11},
  pages   = {115303},
  year    = {2019},
  doi     = {10.1103/PhysRevB.100.115303},
  url     = {https://doi.org/10.1103/PhysRevB.100.115303}
}

@article{wu2021acoustic,
  author  = {Wu, Hong-Wei and Li, Wei and Cheng, Shu-Ling and Yin, Yun-Qiao and Sheng, Zong-Qiang},
  title   = {Acoustic generalized {Kerker} effect},
  journal = {Applied Physics Express},
  volume  = {14},
  number  = {9},
  pages   = {095501},
  year    = {2021},
  doi     = {10.35848/1882-0786/ac1b0e},
  url     = {https://doi.org/10.35848/1882-0786/ac1b0e}
}

@article{Huang2021Sound,
  author  = {Huang, Lujun and Chiang, Yan Kei and Huang, Sibo and Shen, Chen and Deng, Fu and Cheng, Yi and Jia, Bin and Li, Yong and Powell, David A. and Miroshnichenko, Andrey E.},
  title   = {Sound trapping in an open resonator},
  journal = {Nature Communications},
  volume  = {12},
  number  = {1},
  pages   = {4819},
  year    = {2021},
  doi     = {10.1038/s41467-021-25130-4},
  url     = {https://doi.org/10.1038/s41467-021-25130-4}
}

@article{wang2024optical,
  author  = {Wang, Jiajun and Li, Peishen and Zhao, Xingqi and Qian, Zhiyuan and Wang, Xinhao and Wang, Feifan and Zhou, Xinyi and Han, Dezhuan and Peng, Chao and Shi, Lei and Zi, Jian},
  title   = {Optical bound states in the continuum in periodic structures: mechanisms, effects, and applications},
  journal = {Photonics Insights},
  volume  = {3},
  number  = {1},
  pages   = {R01},
  year    = {2024},
  doi     = {10.3788/PI.2024.R01},
  url     = {https://doi.org/10.3788/PI.2024.R01}
}

@article{Deriy2022Compact,
  author  = {Deriy, Ilya and Toftul, Ivan and Petrov, Mihail and Bogdanov, Andrey},
  title   = {Bound States in the Continuum in Compact Acoustic Resonators},
  journal = {Physical Review Letters},
  volume  = {128},
  number  = {8},
  pages   = {084301},
  year    = {2022},
  doi     = {10.1103/PhysRevLett.128.084301},
  url     = {https://doi.org/10.1103/PhysRevLett.128.084301}
}

@article{Huang2022Framework,
  author  = {Huang, Lujun and Jia, Bin and Pilipchuk, Artem S. and Chiang, Yankei and Huang, Sibo and Li, Junfei and Shen, Chen and Bulgakov, Evgeny N. and Deng, Fu and Powell, David A. and Cummer, Steven A. and Li, Yong and Sadreev, Almas F. and Miroshnichenko, Andrey E.},
  title   = {General Framework of Bound States in the Continuum in an Open Acoustic Resonator},
  journal = {Physical Review Applied},
  volume  = {18},
  number  = {5},
  pages   = {054021},
  year    = {2022},
  doi     = {10.1103/PhysRevApplied.18.054021},
  url     = {https://doi.org/10.1103/PhysRevApplied.18.054021}
}

@article{Jia2023ReducedSymmetry,
  author  = {Bin Jia and Lujun Huang and Artem S. Pilipchuk and Sibo Huang and Chen Shen and Almas F. Sadreev and Yong Li and Andrey E. Miroshnichenko},
  title   = {Bound States in the Continuum Protected by Reduced Symmetry of Three-Dimensional Open Acoustic Resonators},
  journal = {Physical Review Applied},
  volume  = {19},
  number  = {5},
  pages   = {054001},
  year    = {2023},
  doi     = {10.1103/PhysRevApplied.19.054001},
  url     = {https://doi.org/10.1103/PhysRevApplied.19.054001}
}

@article{Krasikova2024Helmholtz,
  author  = {Mariia Krasikova and Felix Kronowetter and Sergey Krasikov and Mikhail Kuzmin and Marcus Maeder and Tao Yang and Anton Melnikov and Steffen Marburg and Andrey Bogdanov},
  title   = {Acoustic bound states in the continuum in coupled {Helmholtz} resonators},
  journal = {Physical Review Applied},
  volume  = {22},
  number  = {2},
  pages   = {024045},
  year    = {2024},
  doi     = {10.1103/PhysRevApplied.22.024045},
  url     = {https://doi.org/10.1103/PhysRevApplied.22.024045}
}

@article{Yin2024Rotating,
  author  = {Yuhang Yin and Qilin Duan and Shan Zhu and Jing Li and Zuoti Xie and Cheng-Wei Qiu and Huanyang Chen},
  title   = {Accidental bound states in the continuum in acoustic resonators with rotating obstacles},
  journal = {Physical Review B},
  volume  = {110},
  number  = {5},
  pages   = {054201},
  year    = {2024},
  doi     = {10.1103/PhysRevB.110.054201},
  url     = {https://doi.org/10.1103/PhysRevB.110.054201}
}

@article{MartiSabate2024Observation,
  author  = {Marc Mart{\'i}-Sabat{\'e} and Junfei Li and Bahram Djafari-Rouhani and Steven A. Cummer and Dani Torrent},
  title   = {Observation of two-dimensional acoustic bound states in the continuum},
  journal = {Communications Physics},
  volume  = {7},
  number  = {1},
  pages   = {122},
  year    = {2024},
  doi     = {10.1038/s42005-024-01615-8},
  url     = {https://doi.org/10.1038/s42005-024-01615-8}
}

@article{Huang2024Purcell,
  author  = {Sibo Huang and Shuhuan Xie and He Gao and Tong Hao and Shuang Zhang and Tuo Liu and Yong Li and Jie Zhu},
  title   = {Acoustic {Purcell} effect induced by quasibound state in the continuum},
  journal = {Fundamental Research},
  volume  = {4},
  number  = {1},
  pages   = {57--62},
  year    = {2024},
  doi     = {10.1016/j.fmre.2022.06.009},
  url     = {https://doi.org/10.1016/j.fmre.2022.06.009}
}

@article{Yang2024Merged,
  author  = {Yuzhen Yang and Han Jia and Yunhan Yang and Ping Zhou and Jun Yang},
  title   = {Observation of merged {Fabry-Perot} bound states in the continuum within an underwater phononic crystal},
  journal = {Physical Review B},
  volume  = {110},
  number  = {18},
  pages   = {184108},
  year    = {2024},
  doi     = {10.1103/PhysRevB.110.184108},
  url     = {https://doi.org/10.1103/PhysRevB.110.184108}
}

@article{Yin2024Valley,
  author  = {Shunda Yin and Liping Ye and Hailong He and Xueqin Huang and Manzhu Ke and Weiyin Deng and Jiuyang Lu and Zhengyou Liu},
  title   = {Valley edge states as bound states in the continuum},
  journal = {Science Bulletin},
  volume  = {69},
  number  = {11},
  pages   = {1660--1666},
  year    = {2024},
  doi     = {10.1016/j.scib.2024.04.007},
  url     = {https://doi.org/10.1016/j.scib.2024.04.007}
}

@article{Multibranch2024,
  author  = {Shuowei An and Tuo Liu and Liyun Cao and Zhongming Gu and Haiyan Fan and Yi Zeng and Li Cheng and Jie Zhu and Badreddine Assouar},
  title   = {Multibranch Elastic Bound States in the Continuum},
  journal = {Physical Review Letters},
  volume  = {132},
  number  = {18},
  pages   = {187202},
  year    = {2024},
  doi     = {10.1103/PhysRevLett.132.187202},
  url     = {https://doi.org/10.1103/PhysRevLett.132.187202}
}

@article{Li2025Topology,
  author  = {Weibai Li and Kazem Ghabraie and Xiaodong Huang},
  title   = {Topology-Optimized Bound States in the Continuum with High-{Q} Acoustic Field Enhancement},
  journal = {Advanced Science},
  volume  = {12},
  number  = {21},
  pages   = {2414344},
  year    = {2025},
  doi     = {10.1002/advs.202414344},
  url     = {https://doi.org/10.1002/advs.202414344}
}

@article{Huang2025Collective,
  author  = {Sibo Huang and Shuhuan Xie and Tuo Liu and Tong Hao and Din Ping Tsai and Yong Li and Jie Zhu},
  title   = {Broadband Acoustic {Purcell} Effect from Collective Bound States in the Continuum},
  journal = {Advanced Science},
  volume  = {12},
  number  = {15},
  pages   = {2414627},
  year    = {2025},
  doi     = {10.1002/advs.202414627},
  url     = {https://doi.org/10.1002/advs.202414627}
}

@article{Song2025Sensor,
  author  = {Chao Song and Sibo Huang and Hongyu Ma and Shuhuan Xie and Din Ping Tsai and Jie Zhu and Yong Li},
  title   = {Self-Powered Frequency-Selective Acoustic Sensor Based on Bound States in the Continuum},
  journal = {Advanced Science},
  volume  = {12},
  number  = {15},
  pages   = {2410379},
  year    = {2025},
  doi     = {10.1002/advs.202410379},
  url     = {https://doi.org/10.1002/advs.202410379}
}

@article{Jiao2025Multiband,
  author  = {Junrui Jiao and Dejie Yu},
  title   = {Enhanced multi-band acoustic sensing from quasi-bound states in the continuum},
  journal = {Measurement},
  volume  = {254},
  pages   = {117949},
  year    = {2025},
  doi     = {10.1016/j.measurement.2025.117949},
  url     = {https://doi.org/10.1016/j.measurement.2025.117949}
}

@article{Kronowetter2025Harvesting,
  author  = {Felix Kronowetter and Anton Melnikov and Marcus Maeder and Tao Yang and Yan Kei Chiang and Sebastian Oberst and David A. Powell and Steffen Marburg},
  title   = {Exceptional energy harvesting from coupled bound states},
  journal = {Nature Communications},
  volume  = {16},
  number  = {1},
  pages   = {3515},
  year    = {2025},
  doi     = {10.1038/s41467-025-58831-1},
  url     = {https://doi.org/10.1038/s41467-025-58831-1}
}

@article{BICLines2025,
  author  = {Lin Yang and Riyi Zheng and Sheng Zhang and Wenshuai Zhang and Qiujiao Du and Pai Peng and Ziyu Wang and Manzhu Ke and Xueqin Huang and Fengming Liu},
  title   = {Lines of bound states in the continuum in a phononic crystal slab},
  journal = {Physical Review B},
  volume  = {111},
  number  = {13},
  pages   = {134102},
  year    = {2025},
  doi     = {10.1103/PhysRevB.111.134102},
  url     = {https://doi.org/10.1103/PhysRevB.111.134102}
}

@article{Liu2025Elastic,
  author  = {Feng Liu and Liyun Cao and Wen-Han Guo and Yan-Feng Wang and Yue-Sheng Wang},
  title   = {Elastic bound states in the continuum with multi-polarization hybridization},
  journal = {Communications Physics},
  volume  = {8},
  number  = {1},
  pages   = {392},
  year    = {2025},
  doi     = {10.1038/s42005-025-02301-z},
  url     = {https://doi.org/10.1038/s42005-025-02301-z}
}

@article{He2025Review,
  author  = {Liangshu He and Shengming Sun and Ricardo Martin Abraham-Ekeroth and Yabin Jin and Yanxun Xiang and Dani Torrent},
  title   = {Theoretical developments and experimental Insights of acoustic and elastic bound states in the continuum},
  journal = {npj Acoustics},
  volume  = {1},
  number  = {1},
  pages   = {11},
  year    = {2025},
  doi     = {10.1038/s44384-025-00013-w},
  url     = {https://doi.org/10.1038/s44384-025-00013-w}
}

@article{Hsu2016,
  author  = {Chia Wei Hsu and Bo Zhen and A. Douglas Stone and John D. Joannopoulos and Marin Solja{\v{c}}i{\'c}},
  title   = {Bound states in the continuum},
  journal = {Nature Reviews Materials},
  volume  = {1},
  number  = {9},
  pages   = {16048},
  year    = {2016},
  doi     = {10.1038/natrevmats.2016.48},
  url     = {https://doi.org/10.1038/natrevmats.2016.48}
}

\end{document}